# Designing a Micro-Benchmark Suite to Evaluate gRPC for TensorFlow: Early Experiences

Rajarshi Biswas, Xiaoyi Lu, and Dhabaleswar K. (DK) Panda
{biswas.91,lu.932,panda.2}@osu.edu

**Abstract**

Remote procedure call (RPC) is the backbone of many modern distributed systems. Google's gRPC is one of the most popular open source RPC frameworks available in the community. gRPC is the main communication engine for Google's Deep Learning framework TensorFlow. TensorFlow primarily uses gRPC for communicating tensors and administrative tasks among different processes. Tensor updates during the training phase are communication intensive and thus TensorFlow's performance is heavily dependent on the underlying network and the efficacy of the communication engine. Training deep learning models on TensorFlow can take significant time ranging from several minutes to several hours, even several days. Thus system researchers need to devote a lot of time to understand the impact of communication on the overall performance. Clearly, there is lack of benchmarks available for system researchers. Therefore, we propose TF-gRPC-Bench micro-benchmark suite that enables system researches to quickly understand the impact of the underlying network and communication runtime on deep learning workloads. To achieve this, we first analyze the characteristics of TensorFlow workload over gRPC by training popular deep learning models. Then, we propose three micro-benchmarks that take account these workload characteristics. In addition, we comprehensively evaluate gRPC with TF-gRPC-Bench micro-benchmark suite on different clusters over Ethernet, IPoIB, and RDMA, and present the results.

## 1 Introduction

Deep Learning (DL) a subset of Machine Learning (ML) in Artificial Intelligence (AI) has gotten a lot of attention due to its inference accuracy. Many DL frameworks and tools have been proposed in the community, such as Caffe [17], Facebook Caffe2 [6], Microsoft CNTK [23], Intel BigDL [5], Google TensorFlow [12], and many others.

Google TensorFlow is one of the most popular frameworks to perform distributed deep learning and it has been gaining a lot of momentum recently in Big Data, Deep Learning, and High-Performance Computing (HPC) communities. Distributed TensorFlow uses gRPC [4] as the main communication framework. To achieve high performance tensor transfers, TensorFlow has support for different channels such as native Verbs and MPI. However, these channels still use gRPC for administrative message communication among different remote processes. Therefore, gRPC is an invincible component of TensorFlow. Benchmarking the performance of TensorFlow over gRPC is crucial to identify any possible bottlenecks.

### 1.1 Motivation

The left side of Figure 1 shows the current approach to benchmark deep learning frameworks. Most of the current DL models and benchmarks are deep learning research oriented. In order to get the desired inference accuracy, the Neural Network typically needs a longer training time which makes the benchmarks run longer. Models like VGG [24], AlexNet [18], GoogLeNet [26], etc. take several minutes, or hours, or even days to train on real datasets like ImageNet [3].

However, many system researchers are focused on solely improving the communication engine of deep learning frameworks to reduce the distributed training time. The right side of Figure 1 shows the proposed approach for benchmarking deep learning frameworks for system researchers. They need to consider only factors impacted by the underlying networks and communication subsystem. **Therefore, a micro-benchmark suite that enables system researchers to quickly evaluate, understand, and optimize the performance of the deep learning frameworks' communication substrate in a stand-alone manner by capturing the characteristics of deep learning frameworks is highly desirable.**

Some benchmarks like NCCL2 [10] or Baidu Allreduce [9] aim for reduce tree based collective communication performance evaluation. But, for parameter server based approach, especially for gRPC-based communication runtime, there is a need for micro-benchmarks to evaluate the performance of underlying networks and protocols.

### 1.2 Contribution

To address the above requirements, we propose TF-gRPC-Bench micro-benchmark suite. In this paper, we focus on TensorFlow as the deep learning framework and gRPC as its communication substrate. To achieve our goal, we first characterize TensorFlow workload during training over gRPC communication engine. Then, we design a set of micro-benchmarks by following the distributed TensorFlow communication patterns. In addition, for each of these benchmarks we propose different workload generation schemes that capture distributed TensorFlow's workload patterns over gRPC. Finally, we present the performance results of gRPC using TF-gRPC-Bench miro-benchmark suite over different networks (such as Ethernet, IPoIB) and protocols (such as RDMA).

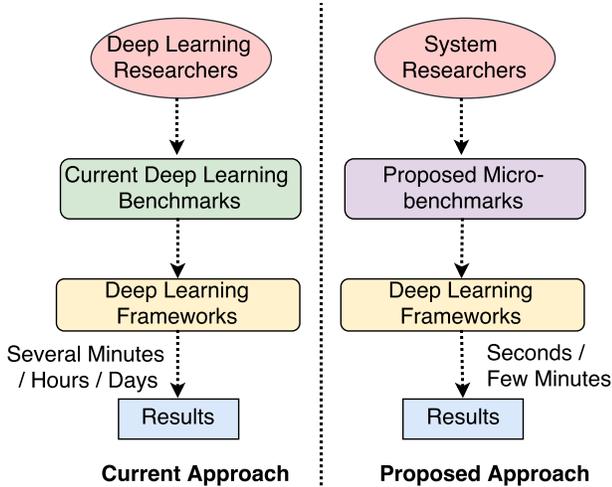

**Figure 1.** Contrast Between Current and Proposed Deep Learning Benchmarks

## 2 Characterization

Before designing the micro-benchmark suite, we first characterize distributed TensorFlow workloads during training of DL models over gRPC.

### 2.1 Distributed execution of TensorFlow

During training, in a distributed TensorFlow cluster values of training variables are updated using aggregated gradients and deltas, represented as tensors. The most widely used approach for managing the training variable in the community is Parameter Server [19]. Figure 2 depicts this communication pattern among TensorFlow parameter servers and workers. In a distributed TensorFlow cluster, the parameter server (PS) processes own the master copies of the variables, whereas, the worker processes request for those variables when needed. In addition, when a worker computes (such as gradient updates) a new value of a variable, it sends an update to the specific PS process. The variable updates also known as the tensor updates are communication intensive and thus the performance is heavily dependent on the underlying networks, protocols and the communication subsystem design.

### 2.2 Tensor Communication over gRPC Channel

Figure 3 depicts tensor communication between a parameter server (PS) and a worker over the gRPC channel. TensorFlow employs rendezvous protocol for tensor transmission among remote processes. In a rendezvous protocol, sender process sends the data only when the receiver process says it is ready. As shown in the figure, the TensorFlow worker (tensor receiving process) actively requests for tensors to the parameter server (tensor sending process). TensorFlow worker issues a RPC call via gRPC library that sends the tensor request to the PS. Upon receiving the request, parameter server finds the requested tensor, and sends to the worker. gRPC core (handles the communication) uses `recvmsg` and `sendmsg` primitives for receiving and sending payloads, respectively. These primitives facilitate receiving/sending data to/from multiple buffers in memory, analogous to scatter-gather operation. These functionalities are achieved with the aid of `iovec` structures. These structures describe the locations and size of each buffer in memory. From now on, in this paper, we refer a buffer pointed by an `iovec` structure as an `iovec` buffer. Figure 3 shows gRPC payload containing multiple such `iovec` buffers. To analyze the characteristics of TensorFlow workload over gRPC channel we examine the `iovec` structures in the gRPC payload in the next section.

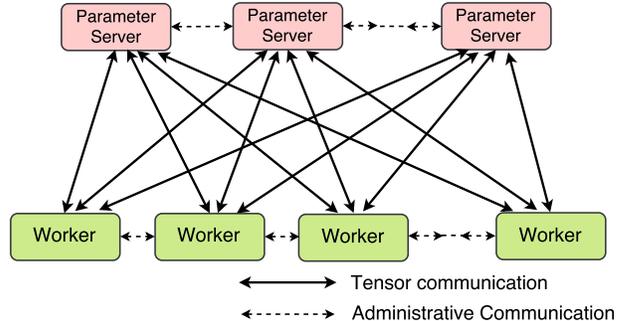

**Figure 2.** Communication Pattern Between TensorFlow Parameter Servers and Workers

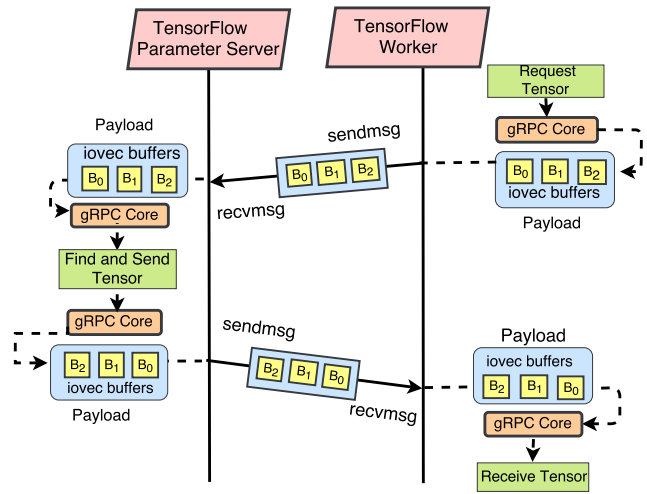

**Figure 3.** Tensor Communication over gRPC Channel Between TensorFlow Parameter Server and Worker

### 2.3 Characteristics of TensorFlow Workload over gRPC Channel

To further understand the pattern of TensorFlow Deep Learning workload over gRPC channel, we profile the `iovec` buffers transmitted during the training phase. For this experiment, we train Resnet [15], VGG [24], Alexnet [18], and Inception [27] Deep Neural Nets (DNN) available in TensorFlow Convolution Neural Network [8] Benchmark on Cluster A. Please see Section 4.1 for specifics about this cluster. We deploy the TensorFlow cluster in Parameter Server mode across five nodes. Two of these nodes act as parameter servers, where as the rest are workers. We use CPUs in parameter servers and GPUs (for compute) in the workers. During training, we kept the batch size 64. Figure 4 shows our observation of the common size distributions pattern of `iovec` buffers in gRPC payloads. In Figure 4 Small, Medium and Large indicate buffers of few Bytes, KBytes and MBytes of length, respectively. As shown in the figure, a gRPC payload may contain a uniform distribution of such Small buffers. On the other hand, a lot of Large buffers and a few Small buffers may create a skew distribution of such buffers in one gRPC payload.

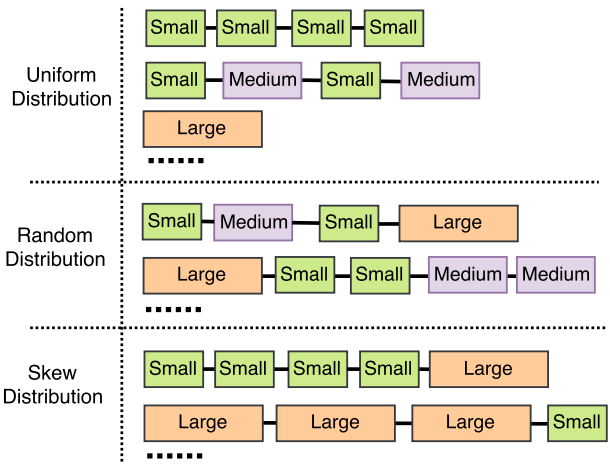

**Figure 4.** `iovec` Buffer Distribution Observed for TensorFlow training over gRPC

## 3 TensorFlow Deep Learning Micro-benchmarks for gRPC

In this section, we first discuss the design considerations for the desired micro-benchmarks. Then, we discuss the actual design of our proposed TF-gRPC-Bench micro-benchmark suite.

### 3.1 Design Considerations

We use our analysis and observations from Section 2 to guide our design considerations for TF-gRPC-Bench benchmarks. We essentially model the distributed TensorFlow communication pattern by only using gRPC. The performance of gRPC can be measured in terms of latency, bandwidth, and throughput. The performance can be significantly influenced by numerous factors such as - number of parameter servers and workers processes, characteristics of the `iovec` buffers, whether data is serialized or not, nature of the underlying network etc. as depicted in Figure 5.

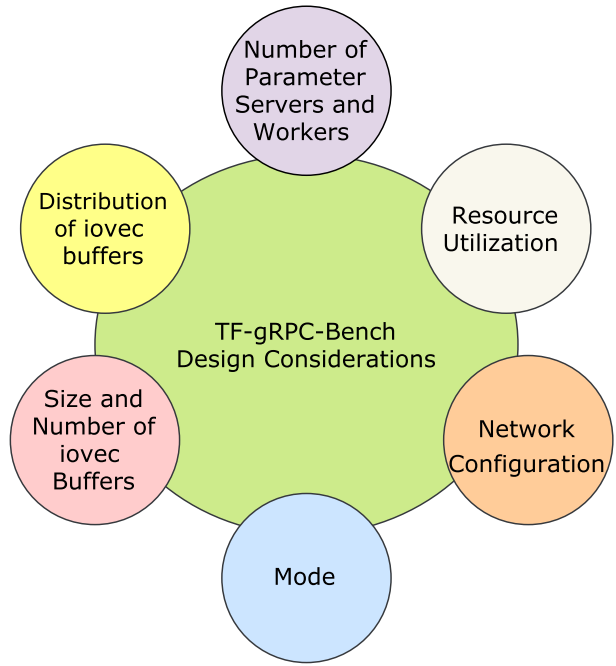

**Figure 5.** Design Considerations for TF-gRPC-Bench Micro-benchmark

Essentially, the efficiency of the network intensive tensor transfers is measured by how fast the parameter servers and workers communicate. Based on these, we consider the following dimensions to design the TF-gRPC-Bench micro-benchmark suite.

**Number of Parameter Servers and Workers** As we have seen in Section 2.1, TensorFlow uses a Parameter Server architecture for distributed training. TensorFlow uses gRPC to communicate (such as tensor transfer) among these remote processes. In such a deployment, the number of parameter servers and workers plays an important role in the overall training performance. For, example deploying only one parameter server for multiple workers may not be efficient. Similarly, deploying more parameter servers may also effect the training time adversely. Therefore, to capture this characteristics, our benchmark models the Parameter Server architecture solely using gRPC and provides flexibility to tune the number of deployed parameter servers and worker processes.

**Distribution of `iovec` buffers** Default gRPC uses `iovec` buffers to construct a payload and uses `recvmsg` and `sendmsg` primitives for communication. Section 2.3 provides insight on the patterns of these buffers distribution in a gRPC payload during TensorFlow training. To capture the characteristics of TensorFlow deep learning workload, we need to consider these patterns to generate payload for the micro-benchmarks.

**Size and number of `iovec` buffers** The size of each individual buffer and the number of such buffers in one payload have a major impact on the performance. For example, if the neural network built using TensorFlow is complex with a large number of input parameters, the tensor updates between parameter servers and workers become increasingly involved. An increase in the tensor size may equate to multiple large buffers in one gRPC payload. Thus controlling these `iovec` buffers in gRPC payload are utmost important for designing the benchmark. In TF-gRPC-Bench, we provide granular control of these buffer size and count to construct the gRPC payload.

**Mode** gRPC uses protocol buffer mechanism for serializing tensors. This serialization can have a constant overhead on gRPC communication engine. Moreover, TensorFlow supports different types of tensor data, hence serialization time may vary which can impact the performance of gRPC. However, to understand the true impact of the underlying network and communication engine on the performance, eliminating serialization overhead is crucial. Therefore, we consider both serialized and non-serialized modes in our micro-benchmark suite design. However, the impact of serialization on RPC frameworks are well studied [16, 20, 22, 25] in the community. Therefore, in this paper, we primarily focus on the non serialized mode.

**Network Configuration** The most crucial operation in distributed TensorFlow training is the tensor updates among different processes. These tensor updates during distributed TensorFlow training result in a many-to-many network intensive communication over gRPC. With large convoluted deep neural network distributed training, the tensor size also increases significantly. Therefore, different network configurations are important parameters to consider, especially when scaling out. This will help the system researchers understand the impact of different networking interconnects and protocols on the distributed training. TF-gRPC-Bench, thus, supports running over any network and cluster configuration.

**Resource Utilization** During the involved Process-to-Process communication of distributed TensorFlow, the gRPC component has a major impact on various computing resources such as CPU, memory, and network etc. With large tensor updates among many parameter servers and workers, the impact on these system resources increases significantly. Therefore, capturing the correlation among different resource utilization while performing network intensive tensor updates over the gRPC channel is essential. Thus, our micro-benchmark suite provides the functionality of measuring different resource utilization during the course of tensor updates.

### 3.2 Design of TF-gRPC-Bench Micro-benchmark Suite

We take the above considerations in account to design our micro-benchmark suite - TF-gRPC-Bench. The design of the benchmark is depicted in Figure 6. Based on the user parameters our benchmark suite first deploys a cluster in Parameter Server architecture to exactly model the distributed TensorFlow communication pattern. We propose three different benchmarks to measure Point-to-Point latency, Point-to-Point Bandwidth and Parameter Server Throughput.

TF-gRPC-Bench supports both serialized and non-serialized mode of payload transfer. For serialized mode, we use gRPC's C++ language binding API's to implement the benchmarks. However, to implement the non serialized mode, we directly use gRPC's core C APIs to avoid any serialization overhead. Table 2 indicates the parameter that can be configured in our benchmark.

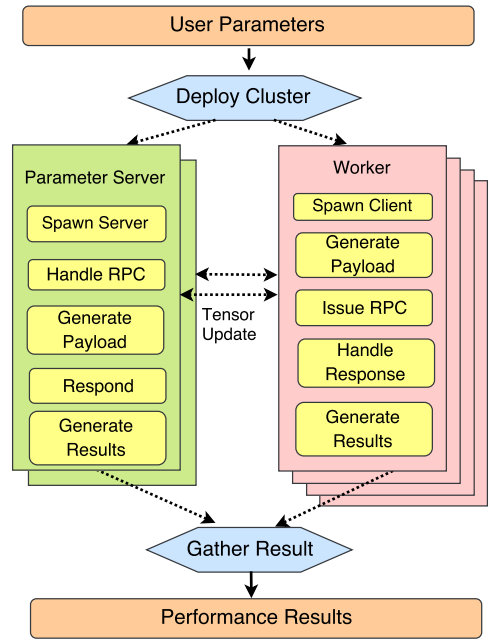

**Figure 6.** TF-gRPC-Bench Micro-benchmark Deign

**TF-gRPC-P2P-Latency** This benchmark measures the Point-to-Point latency of payload transmission between a PS and a worker process. In this benchmark, the RPC procedure in the PS is an echo function that sends back the payload that the worker sends. The processes can be deployed on the same node or different nodes. Users can construct the payload similar to the (we will discuss later in this section in more detail) deep learning workload pattern of TensorFlow over gRPC. In addition to this, users have the flexibility to choose the warm up period, total running period of the benchmark etc. as indicated in Table 2.

**TF-gRPC-P2P-Bandwidth** In this benchmark, we measure the Point-to-Point bandwidth. The worker invokes the remote procedure with a user defined payload and the PS acknowledges the worker request. Similar to the previous benchmark, users have the flexibility to construct payload, defining warm up period, selecting total running time period etc.. This benchmark reports the bandwidth in MBytes per second.

| Category | Default Value | Value Range |
|---|---|---|
| Small | 10 Bytes | [1 Byte - 1 KBytes) |
| Medium | 10 KBytes | [1 KBytes - 1 MBytes) |
| Large | 1MBytes | [1 MBytes - 10 MBytes] |

Table 1. `iovec` Buffer Size Category

**TF-gRPC-PS-Throughput** This benchmark measures the throughput of Parameter Server architecture over gRPC. Users can deploy multiple parameter servers and workers on different processes. This can be run on the same node or different nodes. The performance of the whole system is measured by the aggregated throughput in terms of the number of remote procedures invoked by the workers per second. Each worker invokes remote procedures in all the parameter servers. This is necessary as in distributed TensorFlow, each parameter server is responsible for managing a certain portion of the variables, hence all the workers need to communicate with all the parameter servers to perform updates in the entire variable set. Similar to the previous benchmarks, users have the option to choose payload pattern, warm up period, and total running time of the benchmark etc..

For each of the above benchmarks, the characteristic of the payload is of utmost importance. The essence of TensorFlow deep learning workload must be captured in all the benchmarks payloads. Our benchmark enables users to generate payloads containing Small, Medium and, Large buffers in any pattern. Table 1 presents the acceptable buffer size range of Small, Medium, and Large categories. In addition to the **customized** payload generation, our micro-benchmark suite provides flexibility of automatic payload generation with little or no input from the users. These payloads are generated by taking the observed buffer patterns described in Section 2.3 into consideration. Users have the option to choose any of the payload generation schemes described below.

**Uniform** In this payload generation scheme users can choose to construct the gRPC payload such that `iovec` buffers are distributed uniformly. Users also have the flexibility to choose either Small, Medium, Large buffers or a combination of them in any order. The default value of Small, Medium and Large payloads are 10 Bytes, 10KBytes, 1 Mb. Although, these values are user tunable.

**Random** In this scheme, the buffers are distributed randomly in a gRPC payload. By default, all three buffer categories are used. However, users can choose any type of buffers (at-least two) to automatically generate payload under this scheme.

**Skew** This payload generation scheme distributes the `iovec` buffers unevenly in a gRPC payload. Users need to construct the payload with at-least two different buffer categories. By default, this benchmark chooses all three buffer types. By default, this scheme distributes the buffers specifically keeping biased towards Large buffers because for deep learning workloads Large buffers are more important. For example, if users choose all three categories of the buffers, then one payload will have 60% Large buffers, 30% Medium buffers and 10% Small buffers. Moreover, users have the option to generate the payload in Small or Medium biased manner too.

## 4 Performance Evaluation

In this section, we present comprehensive results of TF-gRPC-Bench micro-benchmark suite. For each of the benchmarks, we use the different data generation schemes available in the micro-benchmark suite and evaluate the performance of gRPC on different clusters and different network interconnects. For workload generation, we use the default configurations as mentioned in Section 3.2, if not specified otherwise. Each payload is constructed using all of the three buffer categories (Table 1) with their default sizes and count (default is total ten buffers in one payload). We use all the three payload generation schemes for distribution of the buffers. Therefore, the skewed distribution generates the largest payload as it contains more Large buffers. We run all the experiments (with default warm up and running time) five times and report the average results. The experiment in section 4.2 use serialized mode, whereas the rest of the experiments use non-serialized mode. In all our experiments, we use gRPC version 1.5.0. For evaluating RDMA, we use our gRPC RDMA [14] design.

### 4.1 Experimental Setup

**(1) Cluster A: Intel Broadwell Cluster (RI2-IB-EDR)**: The RI2 cluster comprises of 20 nodes. Each node is provisioned with Intel Broadwell (E5-2680-v4) dual fourteen-core processors, NVIDIA Tesla K80 GPU, 512 GB of memory, Mellanox IB EDR (100 Gbps) HCA, and a 40G Ethernet. The host processors are running CentOS release 7.2.

**(2) Cluster B: SDSC Comet (SDSC-Comet-IB-FDR)**: The Comet supercomputing system at SDSC has 1,984 compute nodes. Each node is provisioned with Intel Haswell (E5-2680-v3) dual twelve-core processors, 128 GB of memory, 320 GB local SDD, a Mellanox IB FDR (56Gbps) HCA, and 10G Ethernet. The host processors are running CentOS release 6.7.

### 4.2 TF-gRPC-P2P-Latency (Serialized)

First, we evaluate the Point-to-Point latency for gRPC when data serialization mode is enabled on Cluster A. Figure 7 represents the evaluation result for the 64KBytes payload for different communication interconnects. As expected, this figure suggests that the gRPC serialization overhead is constant irrespective of the underlying network. Also, this experiment shows that for 64 KBytes payload Point-to-Point latency is almost similar for both 40G Ethernet and IPoIB on Cluster A. However, RDMA reduces Point-to-Point latency by about 40% compared 40G Ethernet and IPoIB.

| Parameters | Default Value | Value Range | Description |
| --- | --- | --- | --- |
| Benchmark | TF-gRPC-P2P-Latency | TF-gRPC-P2P-Latency, TF-gRPC-P2P-Bandwidth, TF-gRPC-PS-Throughput | Selects a benchmark |
| IP | localhost | Valid IP range | Configures IP of parameter servers |
| Port | 50001 | Valid port range | Configures port of parameter servers |
| Number of parameter servers | 1 | No limit | Controls the number of parameter servers |
| Number of workers | 1 | No limit | Controls the number of workers |
| Mode | Non-Serialized | Non-Serialized/ Serialized | Controls the payload serialization mode |
| Workload generation scheme | Uniform | Uniform, Random, Skew | Generates the payload using the given pattern |
| iovec buffer count | 10 | No limit | Controls the number of iovec buffers in a payload |
| iovec buffers' size | All three categories with default values | Depends on the benchmark | Controls the size of iovec buffers in a payload |
| Warmup time | 2 sec | No limit | Controls warmup seconds for benchmark |
| Total running time | 10 sec | No limit | Controls total running time for benchmark |

Table 2. Configurable Parameters for TF-gRPC-Bench Micro-benchmark Suite

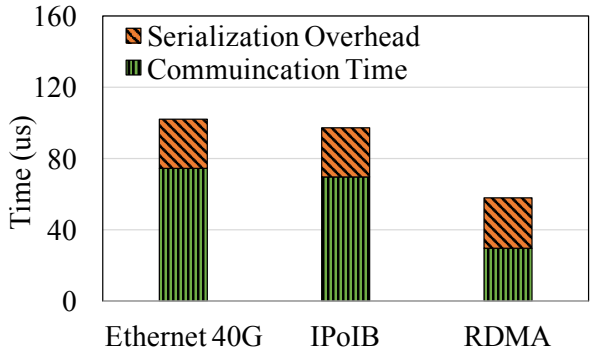

Figure 7. TF-gRPC-P2P-Latency (Serialized) Evaluation on Cluster A with 64KBytes Payload

### 4.3 TF-gRPC-P2P-Latency (Non-serialized)

Next, we evaluate the performance of gRPC in terms of Point-to-Point latency in non-serialized mode. Figure 8 and Figure 9 depict the result of TF-gRPC-P2P-Latency benchmark for different workloads on Cluster A and Cluster B, respectively.

On both clusters, we observe latency is higher for the skewed distribution scheme. This is expected as the default

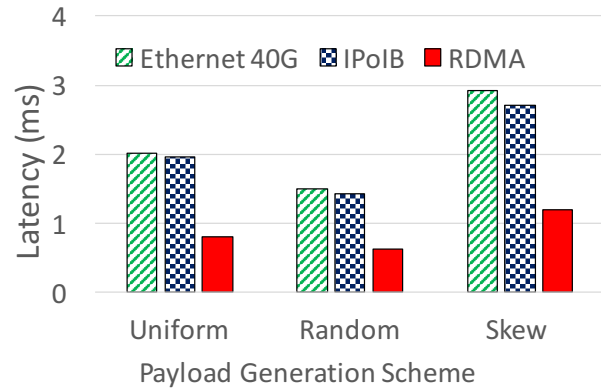

Figure 8. TF-gRPC-P2P-Latency (Non-serialized) Evaluation on Cluster A skewed payload generation scheme is biased towards the iovec Large buffers. Also, we observe that RDMA performs better for all the payload distribution schemes in both the clusters. For example, on cluster A, for skewed payloads RDMA reduces latency by 59% and 56% compared to 40G Ethernet and IPoIB. Similarly, on Cluster B, we observe that RDMA reduces 78% latency compared to 10G Ethernet and 69% compared to IPoIB. In addition, we see IPoIB (FDR-56Gbps) performs almost 27% better than 10G Ethernet on Cluster B.

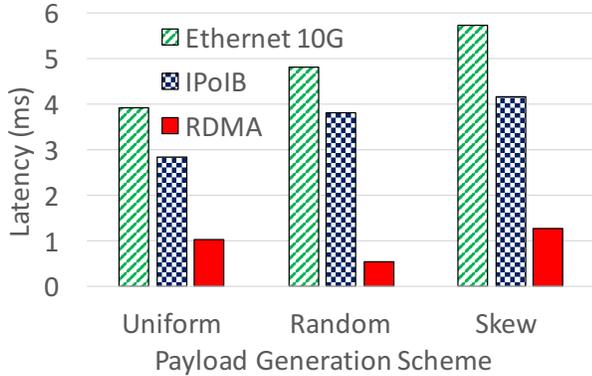

**Figure 9.** TF-gRPC-P2P-Latency (Non-serialized) Evaluation on Cluster B

Moreover, in Figure 10, we compare IPoIB and RDMA gRPC latency on Cluster A with uniformly generated different payloads. We use only Large buffer (1MB each) and vary the number of `iovec` buffer (from two - ten) counts, to generate the payloads. Clearly, RDMA outperforms IPoIB for all payloads. In addition, IPoIB scales poorly with increasing buffer counts in a payload (more total payload size).

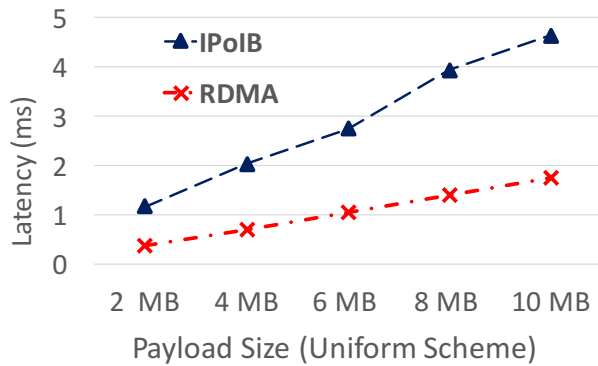

**Figure 10.** TF-gRPC-P2P-Latency (Non-serialized) Evaluation on Cluster A for Different `iovec` Counts

### 4.4 TF-gRPC-P2P-Bandwidth (Non-serialized)

In this section, we present the result of evaluating TF-gRPC-P2P-Bandwidth benchmark. Figure 11 shows the bandwidth obtained in MBytes/second for different payload generation schemes on Cluster A. Similarly, Figure 12 presents the results of running TF-gRPC-P2P-Bandwidth benchmark on Cluster B. On Cluster A, we again observe that gRPC achieves similar bandwidth when the underlying network is either 40G Ethernet or IPoIB. Whereas while using RDMA, gRPC achieves a 2.14x bandwidth compared to IPoIB for skewed distribution scheme. Figure 11 suggests that for random payload generation scheme, IPoIB achieves little less bandwidth than 40G Ethernet. This can be possible for the random nature of the payload generation scheme. As expected, on Cluster B also, RDMA outperforms others. For example, we observe that RDMA achieves 3.2x bandwidth compared to IPoIB for skewed data.

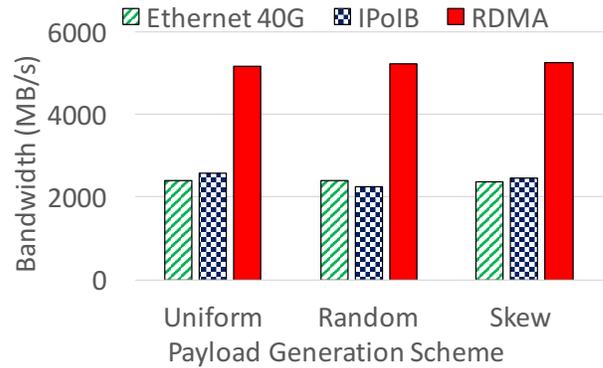

**Figure 11.** TF-gRPC-P2P-Bandwidth (Non-serialized) Evaluation on Cluster A

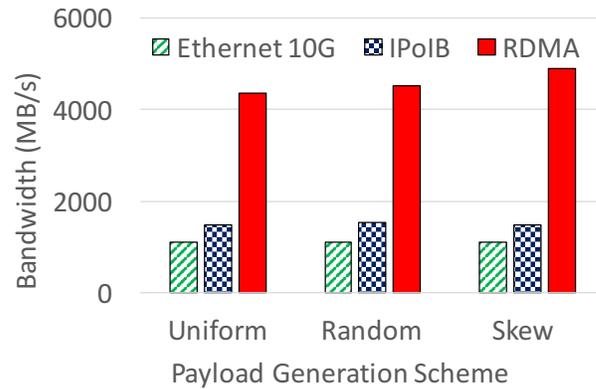

**Figure 12.** TF-gRPC-P2P-Bandwidth (Non-serialized) Evaluation on Cluster B

### 4.5 TF-gRPC-PS-Throughput (Non-serialized)

We show performance of gRPC in terms of throughput in this section. Figure 13 and Figure 14 represent the result of running TF-gRPC-PS-Throughput benchmark with two parameter servers and three workers on Cluster A and Cluster B, respectively. Analyzing the result of this benchmark is of utmost importance, as it essentially mimics TensorFlow communication pattern. The figures show the throughput measured in terms of total RPC calls, invoked by all the workers, per second. Figure 13 indicates that gRPC achieves 4.1x and 3.43x speedup for uniform payload generation scheme when RDMA is used compared to 40G Ethernet and IPoIB on Cluster A. On the other hand on Cluster B, as the Figure 14 suggests, RDMA gRPC achieves better performance than others. For example, gRPC achieves 5.9x speedup with RDMA when compared to 10G Ethernet.

These results indicate that if gRPC uses RDMA, TensorFlow can do the most efficient tensor transfer, hence TensorFlow can achieve low training time. Moreover, using IPoIB compared 40G or 10G Ethernet provides better performance.

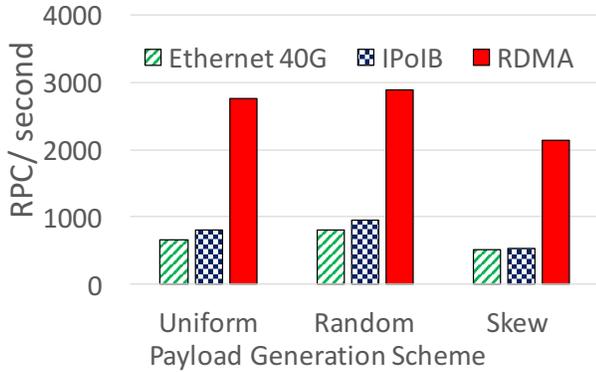

**Figure 13.** TF-gRPC-PS-Throughput (Non-serialized) Evaluation on Cluster A

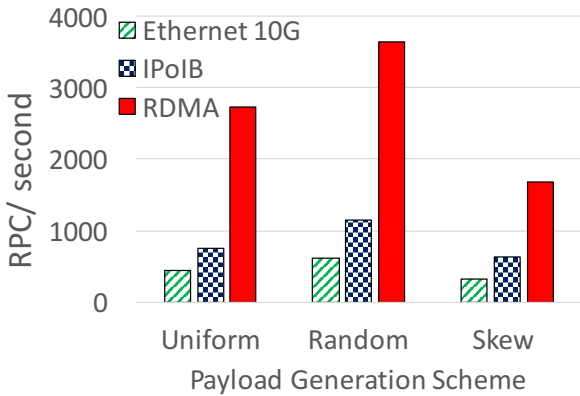

**Figure 14.** TF-gRPC-PS-Throughput (Non-serialized) Evaluation on Cluster B

## 5 Related Work

Remote Procedure Call has come a long way since its inception. The first RPC implementation was presented by Birrell and Nelson [13]. Since then, RPC has evolved and the community has proposed lot of open source high-performance RPC systems such as gRPC [4], Avro [2], and Thrift [1] to name a few. Over the past years, the community has proposed a lot of benchmarks for evaluating RPC frameworks. For example, Lu et al. propose [21] Micro-benchmark suite for Hadoop RPC.

Deep Learning is gaining a lot of attention for its popularity in the Artificial Intelligence domain. As a result benchmarking deep learning applications on different hardware platforms has become important. The official TensorFlow community has support for TensorFlow Convolution Neural Network Benchmark [8] for distributed TensorFlow performance. Baidu research proposes DeepBench [7] primarily to benchmark operations that are important to deep learning on different hardware platforms.

Workload generation is an important part to design efficient deep learning benchmarking. For example, Wang et al. propose BigDataBench [28], a benchmark suite for Big Data Computing, covers typical Internet service workloads and provides representative data sets and data generation tools. The BigDataBench version 4 [11] has support for workload generation for different deep learning frameworks.

To the best of our knowledge, a micro-benchmark suite to benchmark the deep learning communication substrate in a stand-alone manner is not available in the community. In this paper, we propose TF-gRPC-Bench, a micro-benchmark suite for gRPC by taking the characteristics of deep learning workload into account. This benchmark is specially designed for the system researchers' perspective who are solely focused on improving the communication substrate.

## 6 Conclusion and Future Work

gRPC is a widely used Remote Procedure Call framework that enables client and server applications to communicate transparently. This makes easier for the users to create transparent distributed applications and services. gRPC is used as the main communication engine of Deep Learning framework TensorFlow. Nowadays, a majority of the existing clusters are equipped with high performance interconnects such as InfiniBand, 10/25/40/80/100 GigE, RoCE etc.. In order to leverage the benefits of these low latency, high throughput networks, it is evident to study the impact of these networks on the gRPC channel.

In this paper, we propose micro-benchmark suite TF-gRPC-Bench to measure the performance of gRPC over different network interconnects and protocols. We introduce benchmarks to measure Point-to-Point latency, Point-to-Point bandwidth, and Parameter Server throughput that models the distributed TensorFlow communication pattern. Moreover, we analyze the workload characteristics of distributed TensorFlow and design workload for our benchmark that captures the deep learning workload characteristics. TF-gRPC-Bench also provides users the flexibility to configure various parameters including (but not limited to) the payload distribution and size, the number of parameter servers and workers etc.. This benchmark dramatically reduces the experimentation time for the system researchers. Thus, this may help system researchers to quickly evaluate novel communication protocols over different interconnects for deep learning.

As part of our future work, we plan to design micro-benchmarks for other TensorFlow channels such as Verbs, MPI etc. We also intend to make all our benchmarks available through TF-gRPC-Bench package for the community.